\documentstyle[aps,epsfig,multicol,fancyheadings]{revtex}
\begin{document}

\title{Fracture toughness in fibrous materials}

\author{I.\ L.\ Menezes-Sobrinho}

\address{Departamento de F\'{\i}sica, Universidade Federal de Vi\c cosa,
36570-000 Vi\c cosa, MG, Brazil} \date{\today}

\maketitle

\begin{abstract}
In the present paper, a fiber bundle model in (1+1)-dimensions that simulates
the  rupture process of a fibrous material pulled by an uniaxial force
$F$ is analyzed. In this model the load of a broken fiber is shifted in equal
portions onto the nearest unbroken fibers. The force-displacement
diagram is obtained for several traction velocities $v$ and temperatures $t$.
Also, it is shown how the fracture toughness $K_c$ changes with the traction velocity
$v$ and with the temperature $t$.  In this paper it is shown that the rupture
process is strongly dependent on temperature $t$ and on velocity $v$.
\end{abstract}

\noindent {\it PACS \# \ \ 62.20.Mk, 64.60.Fe, 05.40.+j}

\begin{multicols}{2}[]
\narrowtext
\section{Introduction}\label{intro}
Research needs in fracture mechanics are quite varied and still pose a
formidable task for engineers and scientists. When a load sufficiently large is
applied on a material, it fractures in a process that depends on several
factors, such as, the external conditions (temperature, traction
velocity, humidity etc.). The fracture of a material can be classified in two
categories, brittle and ductile \cite{dieter,alman}. These two
categories are not solely functions of the material properties
but depend also on temperature and traction velocity. The brittle fracture
generally occurs at low temperatures and/or high velocities, while the ductile
fracture occurs at high temperatures and/or low velocities.  

When a material is pulled by an
uniaxial force $F$, it experiences a displacement $\delta$. The
force-displacement diagram provides important information about the fracture
process and can be easily obtained experimentally. In this diagram one can
detect a linear region in which the force $F$ increases proportionally to
the displacement $\delta$, obeying Hooke's law. In this region, the
mechanical response of a material is reversible, i.e., if the force is
reset to zero the material returns to exactly the original shape. It is also
observed that, if the force is increased beyond a certain critical value, the
material enters the plastic region, where it does not return to the initial
position when the force vanishes. If the rupture of the material occurs in the
linear region it is called brittle and if the fracture occurs in the plastic
region it is called ductile. Another important information obtained from the
$F$ versus $\delta$ diagram is the fracture toughness, i.e., the amount of
energy needed to fracture the material. The fracture toughness can be
evaluated from the area below the $F$ versus $\delta$ curve. Experimental results
\cite{dieter,smith,tosal} show that the brittle fracture consumes less energy
than the ductile fracture. 

The fracture properties of disordered materials is a subject of great interest
because the presence of disorder is an important feature that
determines the rupture process \cite{livro}. To analyze the rupture process of
disordered materials several models were proposed, among which is the
well-known fiber bundle model (FBM) \cite{zhang,dux,isma1,isma,isma3,ze},
created from the pioneer work of Daniels \cite{dan}. In the FBM a set of fibers
is distributed on a supporting lattice forming a fiber bundle. The fiber bundle
is fixed at both extremes by two parallel plates, one of them is fixed and on
the other an external load is applied.  The FBM can be time independent
(static FBM) \cite{dan,harlow} or not (dynamic FBM)
\cite{isma1,isma,isma3,ze,col,pho,dong}. In the static
model, to each fiber of the bundle is assigned a strength threshold from a
probability distribution and if the applied load exceeds this threshold value
the fiber breaks. In the  dynamic model, each fiber is assumed to have a
lifetime under a given load history, and it breaks because of fatigue. An
important factor in the definition of the FBM is the load-sharing rules, which
describe how the load of a broken fiber is transferred to the unbroken ones. In
equal load sharing (ELS) models the load carried by a broken fiber is equally
distributed among the unbroken fibers of the bundle. In local load sharing
(LLS) the load of a broken fiber is transferred only to its nearest neighbors.
 
In 1994 Bernardes and Moreira introduced an equal load sharing FBM to simulate
fractures in fibrous materials that is sensitive to external conditions,
traction velocity and temperature \cite{ze1}. In this work they obtained
fracture energy (toughness) versus temperature diagrams for
several traction velocities. Then, they concluded that the higher the traction
velocity, the higher is the fracture toughness of the process. These results
indicate that a brittle fracture consumes more energy than a ductile one, in
marked disagreement with the experimental results.  In this paper, a FBM with
local load sharing is studied in order to
analyze the rupture process of a fibrous material pulled by a force $F$ with
a constant velocity $v$. The main goal is to obtain the force-displacement
$F(\delta)$ diagram for several traction velocities $v$ and evaluate the
fracture toughness involved in the rupture process. It is investigated also how
the fracture toughness $K_c$ changes with the traction velocity $v$ and with
the temperature $t$. 
\vfil

\section{Model}\label{model}
The present model was inspired in the one studied by Bernardes and Moreira
\cite{ze1}. It consists of a bundle of
$N_0$ parallel fibers, all with the same elastic constant,
$k$, distributed on a unidimensional lattice.
The fiber bundle is fixed at both extremes by two parallel plates, one of
them is fixed and the other pulled by an uniaxial force $F$ with a
constant velocity $v$. The force $F$ on the fiber bundle is defined as 
\begin{equation}
F=Nk\delta,
\end{equation}
where $\delta$ is the displacement and $N$ is the number of unbroken fibers. 
At each time step the bundle experience an increase $\Delta \delta=v\times
\tau$ in the displacement, where in our units $\tau=1$. In the model
presented here the fiber failure probability depends on the applied load
$\sigma$. The load $(\sigma=F/N)$ is the external force $F$ on the bundle
divided by $N$, the total number of unbroken fibers in the bundle, therefore,
$\sigma=k\delta$. Since the model is of LLS type, an unbroken fiber {\it i}
supports a load $\sigma_i$ given by   
\begin{equation}
\sigma_i=(1+{j\over2})\sigma, \label{sigma} \end{equation}
where $j$ is the number of broken fibers on both sides of the fiber {\it i}.
The failure probability of a fiber $i$ is given by a Weibull distribution
usually used in materials science \cite{livro,shu,wu,zhou} \begin{equation}
P_i(\sigma_i)=1-\exp\left[{-(\sigma_i)^\rho v\over t}\right],
\label{weib} \end{equation}
where $t$ is temperature, $\rho$ is the Weibull modulus, which controls the
degree of disorder in the system, and $v$ is the traction velocity. This
definition of the failure probability is different from that used by Bernardes
and Moreira \cite {ze1} that computed the failure probability from the
elastic energy of a fiber.

 At the beginning of the simulations all fibers are entire and submitted to
the same load $\sigma$ ($j=0$). At each time step fibers are randomly chosen
from a set of $N_q =qN_o$ unbroken fibers. The number $q$ represents a
percentage of fibers and allow us to work with any system size. Then, using
Eqs.~(\ref{sigma}) and (\ref{weib}), the fiber failure probability
$P_i$ is evaluated and compared with a random number $r$ in the interval
[0,1). If  $r<P_i$ the fiber breaks and then the neighboring
unbroken fibers are tested. This procedure describes the propagation of a crack
through the fiber bundle in the direction perpendicular to the
applied force. The process of propagation stops when the test of the
probability does not allow rupture of any other fiber on the border of the
crack or when the crack meets another already formed crack. The same cascade
propagation is attempted by choosing another fiber of the set $N_q$. After all
the $N_q$ fibers have been tested, the bundle is pulled to a new displacement
$\Delta \delta$ and all the rupture process is restarted. The simulation
terminates when all the fibers of the bundle are broken, i.e., when the
bundle is divided into two parts.

\section{Results}\label{scale}
In order to verify the influence of the temperature $t$ and velocity
$v$ in the rupture process of a material, the simulations were
performed considering $N_0=1\times 10^4$ fibers, elastic constant $k=1$, and
Weibull modulus $\rho =2$. The simulations were averaged over 1000
statistically independent samples.

\begin{figure}[f]
\centerline{\epsfig{file=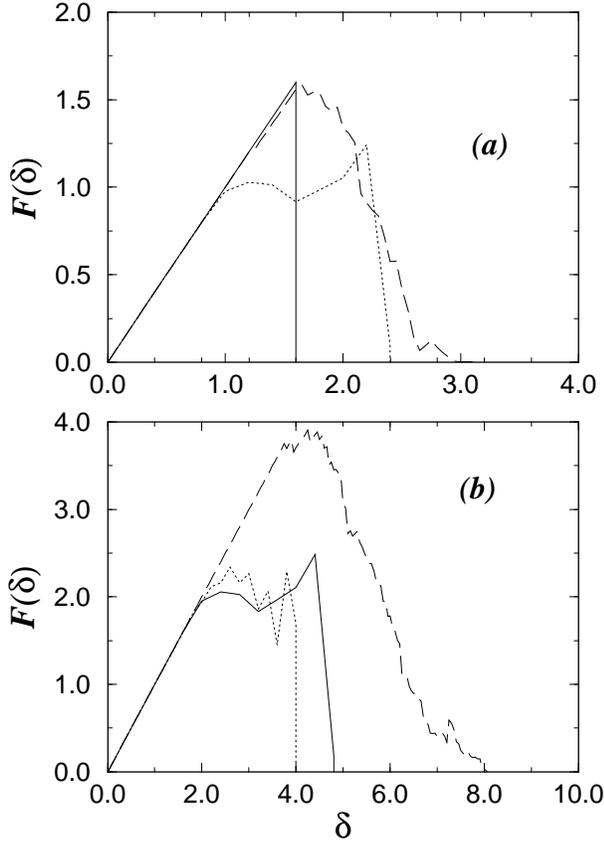,width=8cm,angle=-0}}
\caption{Force $F$ as a function of the displacement $\delta$ for three
different velocities $v$ and two temperatures $t$. In (a) we have $t=0.5$ and 
in (b) $t=4.0$ arbitrary units. $v=0.4$ (solid line), $v=0.2$ (dotted line) and
$v=0.05$ arbitrary units (long dashed).}    \label{tena1}    \end{figure}

Initially, the force-displacement diagrams were obtained in order to verifiy
the influence of temperature $t$ and velocity $v$ on the fracture process.
Figure \ref{tena1} shows the force-displacement diagram $F(\delta)$ for three
velocities $v$ and two different temperatures $t$. In Fig.\ref{tena1} (a) the
results obtained for $t=0.5$ are shown. Note that for $v=0.4$ the relation
between the force $F$ and the displacement $\delta$ is purely linear.  This
behavior is characteristic of a brittle fracture, where the rupture occurs
due to the appearance of big cracks in the material. For low and intermediate
velocities the relation between the force $F$ and the displacement $\delta$ is
not purely linear and in this case the fracture occurs in the brittle-ductile
transition or in the ductile region. From Fig.\ref{tena1} (a) one can see
that the lower the velocity, the greater the area below the
force-displacement curve.  

In Fig.\ref{tena1} (b) the results were obtained
for a temperature $t=4.0$. Now, for any velocity $v$, the displacement diagram
is not purely linear, i.e., for this temperature and these sets of
velocities the fracture will not be brittle. In order to better understand the
influence of the temperature $t$ and velocity $v$ on fracture process, we will
discuss how the fracture toughness $K_c$ is influenced by the temperature $t$
and the velocity $v$.

\begin{figure}[f]
\centerline{\epsfig{file=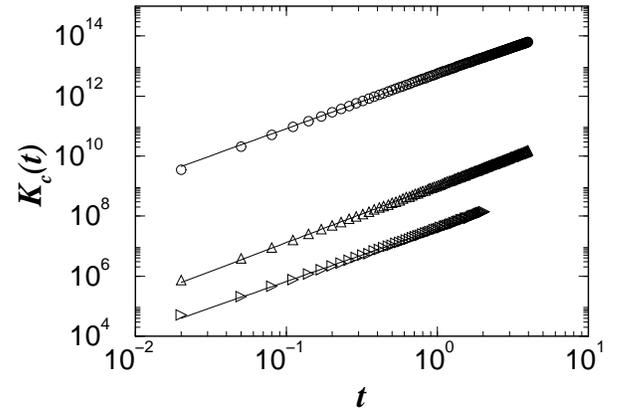,width=8cm,angle=-0}}
\caption{Log-Log plot of the fracture toughness $K_c$ as a function of the temperature
$t$ for three different velocities: $v=0.002$ (circles), $v=0.02$ (up
triangles) and $v=0.05$ (right triangles).}  \label{tena2} 
\end{figure}

\begin{figure}[f]
\centerline{\epsfig{file=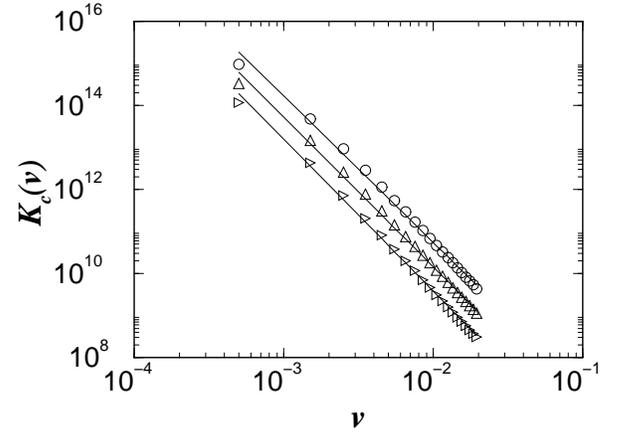,width=8cm,angle=-0}}
\caption{Log-Log plot of the fracture toughness $K_c$ versus the velocity
$v$ for three different temperatures: $t=2.0$ (circles), $t=1.0$ (up
triangles) and $t=0.5$ (right triangles).}  \label{tena3} 
\end{figure}

Figure \ref{tena2} shows a log-log plot of the fracture toughness $K_c$ as a function of
the temperature $t$ for three different velocities, $v=0.002$, $v=0.02$,
and $v=0.05$. Note that the fracture toughness $K_c$ increases linearly  with the
increase of the temperature $t$, indicating a power law 
\begin{equation}
K_c\simeq t^\alpha,
\label{wei} \end{equation} 
where $\alpha$ is an exponent that depends on the velocity. So, the higher the
temperature $t$, the more energy will be absorbed before a catastrophic rupture
occurs. Also, Fig. \ref{tena2}(a) shows that the higher the velocity $v$,
the lower the fracture toughness $K_c$, i.e., a smaller quantity of
energy will be spent in the fracture process. In Fig. \ref{tena3} this fact
is shown more clearly. It shows a log-log diagram of the fracture toughness
$K_c$ versus the velocity $v$ for three temperatures. Note that the fracture
toughness decreases with the increase of the velocity. 

The results obtained agree with experimental data obtained in fracture
mechanics \cite {dieter,smith,tosal}. It is well known that the failure of
materials has a strong dependence on temperature and velocity, for example, as
is the case for the failure behavior of polymers. At
very low temperatures they fracture brittle and consume little energy during
the rupture process \cite{dieter,livro,ward}. When the temperature is
increased above of the  critical temperature $t_c$, the polymer undergoes a
transition to rubber-like behavior in which the material can be elastically
stretched over several times its initial size. In this region the fracture
process is slow and consumes very much energy. Also, the rupture behavior of
polymers is strongly dependent on the speed in which the elongation takes
place.                           

\section{Conclusion}\label{discussion}
In conclusion, we studied a model for fracture on fibrous materials in
(1+1) dimensions that simulates a rupture process sensitive to temperature
$t$ and to velocity $v$. It is well known that the fracture toughness,
i.e., the energy (work) consumed to break the material is
strongly dependent on temperature and on traction velocity. At low temperatures
and/or high velocities the fracture toughness is lower than that in high
temperatures and/or low velocities. 

In Ref. \cite {ze1} Bernardes and Moreira, used a fiber bundle model for
which the fracture toughness is sensitive to temperature $t$ and to velocity $v$.
However, their results do not agree with experimental
observations. In the present paper, it was studied a similar model to the
one used by Bernades and Moreira \cite {ze1} and the force-displacement
diagrams for three different velocities and two temperatures were obtained. In
these diagrams, one can observe two regions dependent on temperature $t$ and
velocity $v$, an elastic and a plastic region. In the elastic region the force
$F$ is proportional  to displacement $\delta$. In the plastic region the force
$F$ is not linearly proportional to the displacement $\delta$
and with the increase in $\delta$ it reaches a maximum value, beyond which it
decreases. The area below the force-displacement curve give us the toughness
$K_c$ and depends on the temperature and velocity. The results obtained in
this work show that the fracture toughness $K_c$ increases with the increase
of the temperature $t$ and decreases with the increase of the velocity $v$.
These results are in agreement with the experimental observations.

\medskip
\centerline{\bf Acknowledgments}
\medskip

\noindent
The author thanks M. L. Martins, M. S. Couto, and J. A. Redinz for helpful
criticism of the manuscript and the kind hospitality of the Departamento de
F\'{\i}sica, UFMG. This work was supported by FAPEMIG (Brazilian agency).   


\end{multicols}
\widetext
\end{document}